\documentstyle[aps,preprint,prl]{revtex}
 \tighten
\begin{document}
\draft
\title{Simple Realization Of The Fredkin Gate Using A Series Of Two-body
 Operators\footnotemark{}}
\footnotetext{$^*$\ Accepted by {\it \prl}}
\author{H. F. Chau and F. Wilczek}
\address{
 School of Natural Sciences, Institute for Advanced Study, Olden Lane,
 Princeton, NJ 08540, U.S.A.
}
\date{\today}
\preprint{IASSNS-HEP-95/15; quant-ph:9503005}
\maketitle
\begin{abstract}
 The Fredkin three-bit gate is universal for computational logic, and is
 reversible. Classically, it is impossible to do universal computation using
 reversible two-bit gates only. Here we construct the Fredkin gate using a
 combination of six two-body reversible (quantum) operators.
\end{abstract}
\medskip
 Since the pioneering work of Feynman and Deutsch \cite{Foundation,Feynman},
 the potential to do universal computation in a closed system, using elements
 following the laws of quantum mechanics, has been recognized. There are also
 important problems for which it seems likely that quantum computers, if they
 can be realized, will have capabilities qualitatively superior to classical
 ones. The most obvious problems in this class, perhaps, involve simulation of
 the dynamics of quantum systems. Recently Shor \cite{Shor} discovered a much
 less obvious application to a naturally defined problem: factorizing large
 numbers. His probabilistic algorithm for factorizing large composite numbers
 $N$ using a quantum computer whose running time is polynomial in $\log N$,
 whereas all known classical algorithms are non-polynomial in this variable.
 Other relevant investigations include explorations of the possible physical
 implementation of quantum computers \cite{Optical_Fredkin,Impl}, quantum
 computational complexity classes \cite{Complexity}, quantum teleportation
 \cite{Teleportation}, and quantum coding \cite{Coding}.
\par
 In the earliest work, the question of how a quantum mechanical computer, whose
 operation relies on (reversible) unitary matrices, can perform classical
 irreversible logical operations like AND was addressed. It was realized that
 reversibility can be maintained at the price of carrying around extra
 ``garbage bits''. Indeed, previous work on classical reversible computation
 had demonstrated that one could construct a universal machine using simple
 reversible three-in three-out prototypes. In particular the Fredkin gate
 \cite{FredkinGate} (see Fig.~\ref{F:FredGate}), whose characteristic is
 tabulated in Table~\ref{T:Truth} is known to be universal. For example, by
 fixing $c_i = 0$ in the input, it is easy to verify that $c_o$ gives us the
 logical AND between $a_i$ and $b_i$ in the output. Other irreversible logical
 operations can be recovered in a similar manner. We represent 0 and 1 by
 $|0\rangle$ and $|1\rangle$ respectively. Then quantum mechanically, the
 Fredkin gate logic corresponds to the following three-body unitary
 transformation:
\begin{equation}
 U_{\rm Fredkin} = I + a^{\dag} a \left( b^{\dag} c + c^{\dag} b - b^{\dag} b
 - c^{\dag} c + 2 b^{\dag} b c^{\dag} c \right) \mbox{,} \label{E:U_Fredkin}
\end{equation}
 where $a$ and $a^{\dag}$ denote the annihilation and creation operators at
 site $a$ respectively. Since $U_{\rm Fredkin}$ is a three-body operator, its
 direct implementation would seem to require delicate cancellation of more
 fundamental two-body interactions, to leave behind a specific complicated
 three-body residual, which is very awkward.
\par
 Thus one is motivated to inquire whether $U_{\rm Fredkin}$ can be constructed
 using a (finite) composition of two-body operators. This question has been
 partially answered by DiVincenzo \cite{Two-bit}, who proposes a method to
 approximate Fredkin gate logic up to any accuracy $\epsilon > 0$ by
 $O(1/\sqrt{\epsilon})$ two-body unitary operators. Clearly, this result leaves
 room for improvement. DiVincenzo and Smolin \cite{N_Two-bit} have done
 extensive numerical work, producing convincing evidence that any three-bit
 gate can be constructed by a suitable combination of six two-bit gates. In
 this note, we explicitly construct $U_{\rm Fredkin}$ using six two-body
 unitary operators. By way of contrast no combination of classical reversible
 two-bit gates is sufficient for universal computation, so that our
 construction provides another example of a qualitative enhancement of
 computational power through quantum mechanics.
\par
 Let us now introduce three basic gates used in our construction. A quantum-NOT
 gate is a one-in one-out quantum gate (see Fig.~\ref{F:Cond_U_Gate}(a)),
 performing the unitary transformation
\begin{equation}
 N_{(a)} |\alpha\rangle_a = \sigma_1 |\alpha\rangle_a \mbox{,} \label{E:Not}
\end{equation}
 where $\sigma_i$ are the Pauli spin matrices, is an extension of the classical
 logical NOT to the quantum regime. We may also interpret the quantum-NOT gate
 as the one which gives $(a+1) \mbox{~mod~} 2$ from an input q-bit $a$.
 Diagrammatically, we represent a quantum-NOT gate as a rectangular box labeled
 by $N$ (see Fig.~\ref{F:Cond_U_Gate}(a)). Clearly, this is a one-body
 operator.
\par
 A conditional-$U$ gate is a two-in two-out quantum logic gate (see
 Fig.~\ref{F:Cond_U_Gate}(b)), performing the unitary transformation
\begin{equation}
 U_{(a,b)} |\alpha\rangle_a |\beta\rangle_b = \left( 1 - a^{\dag} a \right)
 |\alpha\rangle_a |\beta\rangle_b + a^{\dag} a |\alpha\rangle_a
 U|\beta\rangle_b \mbox{.} \label{E:Cond_U}
\end{equation}
 Here, $|\alpha\rangle_a$ is used as a control, whose state will not change
 after passing through the gate. When $|\alpha\rangle_a = |0\rangle$, the gate
 does nothing. And when $|\alpha\rangle_a = |1\rangle$, state $|\beta\rangle_b$
 is mapped to $U|\beta\rangle_b$. As shown in Fig.~\ref{F:Cond_U_Gate}(b), we
 represent a conditional-$U$ gate by a rectangular box labeled by $U$. The
 control q-bit ($a$ in this case) is represented by drawing a dash line between
 the input ($a_i$) and the output ($a_o$). A conditional-$U$ gate defines a
 two-body operator.
\par
 In particular, the conditional-$\sigma_1$ is of great importance. One can
 write down the ``truth table'' of this gate and find that it performs
 conditional NOT on the second q-bit $b$ using the first q-bit $a$ as control.
 In order to make the meaning of this gate more apparent, we denote this gate
 by conditional-$N$.
\par
 Finally, we introduce a doubly-controlled phase shifter (see
 Fig.~\ref{F:Cond_U_Gate}(c)), which performs
\begin{equation}
 P_{(a,b)} |\alpha\rangle_a |\beta\rangle_b = |\alpha\rangle_a |\beta\rangle_b
 - ( 1 - P ) a^{\dag} a |\alpha\rangle_a b^{\dag} b |\beta\rangle_b \mbox{,}
 \label{E:Double_Cond_Phase}
\end{equation}
 for some phase rotation $P = e^{i\theta}$. This is again a two-body operator,
 which changes the phase of the overall wavefunction provided that both $a$ and
 $b$ are in state $|1\rangle$, while q-bit $c$ is entirely passive. We
 represent a doubly-controlled phase shifter as by a rectangular box labeled by
 $P$ (see Fig.~\ref{F:Cond_U_Gate}(c)).
\par
 We can now record our Fredkin gate construction. As shown in
 Fig.~\ref{F:Construction}, it is a four stage construction consequentially
 making up of an adder, an $i\sigma_1$ generator, an $i$ remover, and a
 subtracter. It corresponds to the following equation:
\begin{eqnarray}
 U_{\rm Fredkin} (a,b,c) & = & N_{(c,b)} \circ N_{(c)} \circ P_{(a,b)} \circ
 U_{(a,c)} \circ V_{(b,c)} \circ
 U_{(a,c)} \circ V_{(b,c)}
 \circ N_{(c)} \circ N_{(c,b)} \mbox{,} \label{E:Result}
\end{eqnarray}
 where $U = \sigma_2$, $V = (\sigma_2 + \sigma_3) / \sqrt{2}$, and $P = -i$.
 Since the first two of these operators act only on q-bits $b$ and $c$, they
 can be combined; similarly the last three can be combined. Thus we have a six
 two-body gate realization as advertised.
\par
 Let us now explain how this construction works. First we want to gather all
 the quantum states that might be changed upon passing through a Fredkin gate
 to the third q-bit $c$. The most economical way to do this is by performing an
 addition modulo 4 in q-bits $b$ and $c$. This can be done by a combination of
 a quantum-NOT and a conditional-$U$ gates (see Fig.~\ref{F:Construction}).
 After the core computation, we can of course reverse the above process using a
 subtracter. This accounts for a total of four gates. Using $|0,0,0\rangle$,
 $|0,0,1\rangle$, $\cdots$, $|1,1,1\rangle$ as our basis, and denoting them by
 $1$, $2$, $\cdots$, $8$ respectively, then the combined action of the adder
 and the subtracter is to relabel the basis in the order of $4$, $1$, $2$, $3$,
 $8$, $5$, $6$, and $7$ . In the new representation, $U_{\rm Fredkin}$ becomes
 the matrix
\begin{equation}
 U_{\rm Toffoli} = \left[ \begin{array}{cc} I_6 & 0 \\ 0 & \sigma_1 \end{array}
 \right] \mbox{,} \label{E:U_Toffoli}
\end{equation}
 where $I_6$ is the $6\times 6$ identity matrix. The $U_{\rm Toffoli}$ logic,
 which is sometimes called the Toffoli gate \cite{ToffoliGate} or the
 ``controlled controlled  NOT gate'' \cite{Feynman}, is also known to be
 universal. The convenient feature of the new basis is that the first two
 q-bits $a$ and $b$ are unaltered after the operation $U_{\rm Toffoli}$. In
 addition, the third q-bit $c$ changes its state when and only when $a$ and $b$
 are both spin up.
\par
 Inspired by the idea of commutators in group theory, we ask if it is possible
 to construct two conditional-$U$ gates such that
\begin{equation}
 U_{(a,c)} \circ V_{(b,c)} \circ U^{-1}_{(a,c)} \circ V^{-1}_{(b,c)} =
 U_{\rm Toffoli} \mbox{.} \label{E:UVU-1V-1}
\end{equation}
 This is possible when
\begin{equation}
 U V U^{-1} V^{-1} = \sigma_1 \mbox{.} \label{E:UV}
\end{equation}
 Unfortunately, Eq.~(\ref{E:UV}) cannot be satisfied. A contradiction is
 arrived by taking the determinant in both sides of the equation. However, if
 we replace $\sigma_1$ by $i\sigma_1$ in Eq.~(\ref{E:UV}), like what DiVincenzo
 has done in Ref~\cite{Two-bit}, solutions can be found. One of the possible
 solutions is $U = \sigma_2$ and $V = (\sigma_2 + \sigma_3) / \sqrt{2}$. This
 solution has a nice feature that $U = U^{-1}$ and $V = V^{-1}$, which makes
 the actual construction of the machine a bit simpler. We call it the
 $i\sigma_1$ generator in Fig.~\ref{F:Construction}, which eats up another four
 two-body gates. One can show that it is a minimal construction, in the sense
 that any proposal involving fewer than four two-bit gates cannot do the same
 computation. Alternative constructions of the Toffoli gate has been proposed
 by various authors \cite{Toffoli_Construct}.
\par
 Finally, we have to remove the extra phase $i$ from the system. This can be
 done trivially by using a doubly-controlled phase shifter with $P = -i$ (see
 Fig~\ref{F:Construction}). This completes our construction.
\par
 In summary, we have explicitly constructed a sequence of six two-body quantum
 gates to realize the three-in three-out Fredkin gate logic. As we have
 mentioned, this bypasses one significant barrier toward the possible
 construction of a quantum computer. Our construction can be used in the
 realization of other similar quantum gates as well. For example, the matrix
\begin{equation}
 M = \left[ \begin{array}{cc} I_6 & 0 \\ 0 & \cos \lambda + i \sin \lambda
 \sigma_1 \end{array} \right] \mbox{,} \label{E:Further}
\end{equation}
 which appears in Eq.~(3.4) of Ref~\cite{Two-bit}, can be simulated by choosing
 $U = \sigma_2$ and $V = \cos ( \lambda / 2 ) \sigma_2 + \sin ( \lambda / 2 )
 \sigma_3$. By replacing this set of $U$ and $V$ in Fig.~\ref{F:Construction},
 a generalized quantum Fredkin gate is obtained. Details of other efficient
 quantum logic gate constructions will be reported elsewhere \cite{Continue}.
\par
 It would be interesting to know if the Fredkin gate can be built using fewer
 than six quantum two-body gates. If we only demand the output of a quantum
 Fredkin gate to be correct up to a phase, then Milburn \cite{Optical_Fredkin}
 provides a three gate construction, but this is not a suitable building block
 for universal quantum computation. We believe that a construction of a true
 Fredkin gate using fewer than six quantum two-body gates, if possible, would
 have to involve a substantially different idea.
\acknowledgments{This work is supported by DOE grant DE-FG02-90ER40542.}

\begin{table}
 \begin{tabular}{ccc|ccc}
  \multicolumn{3}{c|}{Input} & \multicolumn{3}{c}{Output} \\
  $a_i$ & $b_i$ & $c_i$ & $a_o$ & $b_o$ & $c_o$ \\ \hline
  0 & 0 & 0 & 0 & 0 & 0 \\
  0 & 0 & 1 & 0 & 0 & 1 \\
  0 & 1 & 0 & 0 & 1 & 0 \\
  0 & 1 & 1 & 0 & 1 & 1 \\
  1 & 0 & 0 & 1 & 0 & 0 \\
  1 & 0 & 1 & 1 & 1 & 0 \\
  1 & 1 & 0 & 1 & 0 & 1 \\
  1 & 1 & 1 & 1 & 1 & 1
 \end{tabular}
 \caption{The ``truth table'' of a Fredkin gate.}
 \label{T:Truth}
\end{table}
\begin{figure}
 \caption{Fredkin gate, $a_i$, $b_i$, $c_i$ are the inputs, while $a_o$, $b_o$,
  $c_o$ are its outputs.}
 \label{F:FredGate}
\end{figure}
\begin{figure}
 \caption{(a) a quantum-NOT gate; (b) a conditional-$U$ gate; and (c) a
  doubly-controlled phase shifter. We represent the control bit by drawing a
  dash line between its input and output.}
 \label{F:Cond_U_Gate}
\end{figure}
\begin{figure}
 \caption{Construction of Fredkin gate using two one-body and seven two-body
  quantum gates.}
 \label{F:Construction}
\end{figure}
\end{document}